\documentclass[12pt,a4paper]{article}

\usepackage{amsmath,amssymb,amsfonts,amsthm}

\usepackage{t1enc}
\usepackage[english]{babel}
\usepackage[latin1]{inputenc}
\usepackage{color}
\usepackage{graphicx}

\usepackage{hyperref}

\theoremstyle{definition}

\newcommand{\ra}{\rightarrow}

\title{CMC surfaces and area-charge inequality for a spheroidal ECD spacetime}
\author{Andr\'es Ace\~na\\ 
\\  
  Facultad de Ciencias Exactas y Naturales,\\ Universidad Nacional de Cuyo, \\ Instituto Interdisciplinario de Ciencias B\'asicas, CONICET, \\ Mendoza, Argentina.
  }
\date{}

\begin{document}
\maketitle

\begin{abstract}
 We consider the spacetime presented by Bonnor \cite{Bonnor98}, whose matter content is a spheroid of electrically counterpoised dust, in the context of the geometrical inequalities between area and charge. We determine numerically the constant mean curvature surfaces that are candidates to be stable isoperimetric surfaces and analyze the relation between area and charge for them, showing that a previously proved inequality is far from being saturated. We also show that the maximal initial data has a cylindrical limit where the minimum of the area-charge relation is attained.
\end{abstract}

\section{Introduction}\label{sec1}

General relativity, being fundamentally a geometric theory, relies heavily in geometrically defined objects to make physical predictions. In fact, many theoretical predictions depend on defining suitable geometric objects associated with physical concepts and then prove geometrical relations among them. As an already classical and crucial example we have the positivity of mass, $m\geq0$, for isolated systems \cite{Witten81}, \cite{ShoenYau81}. An important family of such relations is formed by the so called geometrical inequalities. They are predictions of the theory and allow us to understand relevant properties of physical systems. For black holes they tend to relate the area of the black hole and physical parameters, such as the mass, charge and angular momentum. As example we have that for axially symmetric black holes
\begin{equation}
 \sqrt{|J|}\leq m,
\end{equation}
proved in \cite{Dain08}, where $J$ is the angular momentum of the black hole. Also in this category we have the Penrose inequality \cite{Penrose73},
\begin{equation}
 m\geq\sqrt{\frac{A}{16\pi}},
\end{equation}
which presents still a very important open problem. Crucially this last inequality and part of the motivation for the former (and other inequalities) depend on the cosmic censorship conjecture. For an up to date and thorough review of geometrical inequalities in general relativity the reader is referred to \cite{DainGabachClement18} and for the present status of the Penrose inequality to \cite{Mars09}.

A problem that turned out to be extremely challenging is to find geometrical inequalities for ``ordinary objects'', that is, regular matter objects. One of the reasons for such difficulty is that ordinary objects do not have geometrically distinguished boundaries as black holes do. After the firsts struggles to find geometric relations or conditions on matter objects it was realized that simple attempts to relate things like the mass and area or length of an object were bound to fail unless spherical symmetry was present. In this sense it became clear that just the length of an object or its area was not a good measure of its ``size'', as it was possible to make the area of an object go to zero without making its mass zero, or it was possible to make an arbitrarily long object with fixed mass. A much used example of this, an the main concern of the present article, is the spacetime presented in \cite{Bonnor98}. Progress was made once it was realized that an extremality condition was lacking, which was something that the black holes naturally had. To the rescue came trapped surfaces for the spacetime and isoperimetric surfaces for initial data. A stable isoperimetric surface is a surface whose area is a minimum  with respect to nearby surfaces that enclose the same volume. In this sense, the area of the smallest stable isoperimetric surface enclosing an object is seen as a better representative of the size of the object than the area of the object itself. Please refer again to \cite{DainGabachClement18} for the state of the art and the discussion of possible measures of size for ordinary objects.

The problem of finding the stable isoperimetric surfaces in a given metric manifold, so called isoperimetric problem, has a long and fruitful history in mathematics. Of particular importance in general relativity are the results showing that for initial data that is asymptotic to Schwarzschild there exists a unique foliation by stable isoperimetric surfaces in the asymptotic region. In this regard please refer to \cite{HuiskenYau96}, \cite{EichmairMetzger2013a} and \cite{EichmairMetzger2013b} (this last work contains also an interesting appendix with an overview of results on isoperimetric regions).

Back to geometrical inequalities, it was proved  in \cite{DainJaramilloReiris21} that for a stable isoperimetric surface in an electro-vacuum maximal initial data with a non-negative cosmological constant
\begin{equation}\label{desProbada}
 A\geq \frac{4}{3}\pi Q^2,
\end{equation}
where $A$ is the area of the surface and $Q$ is its electric charge. This result is purely quasi-local and therefore charged matter could be present inside or outside the surface. To explore this inequality in \cite{AcenaDain13} the super-extreme regime of the Reissner-Nordstr\"om family of spacetimes was considered. Due to the staticity and spherical symmetry the isoperimetric profile was obtained and then it was possible to see that equality in \eqref{desProbada} was never attained. This led to conjecture the more stringent inequality
\begin{equation}\label{desConjetura}
 A \geq \frac{16}{9}\pi Q^2.
\end{equation}

Continuing with the idea of testing the inequality \eqref{desProbada}, in the present work we focus on the family of spacetimes presented by Bonnor in \cite{Bonnor98}. These are solutions of the Einstein-Maxwell field equations where the matter content is a spheroid of electrically counterpoised dust (ECD). These solutions are striking in the way the mathematical properties of ECD are exploited. This allows to modify the parameters of the spacetimes in order for example to test the hoop conjecture \cite{Thorne72}. Also, the area of the matter spheroid can be made arbitrarily small while keeping its mass (and charge) constant. This is a counterexample to the idea that the area of an object can be bounded by its mass.  Our main interest is to analyze these spacetimes in relation to the inequality \eqref{desProbada}. So we want to find the isoperimetric profile for a maximal initial slice and obtain the quotient between area and charge. From the perspective of  the conjecture \eqref{desConjetura}, we are interested in seeing if the spacetime presents a challenge to it.

To find the stable isoperimetric surfaces in a metric manifold is in general quite complicated. A necessary condition for a surface to be stable isoperimetric is that its mean extrinsic curvature has to be constant \cite{BarbosaDoCarmoEschenburg88}. Such a surface is said to be of constant mean curvature (CMC). Even finding CMC surfaces is in general complicated, and it is necessary to use algorithms as the one developed in \cite{Metzger04}. We have the benefit that the spacetime we consider is static and axially symmetric, which allows us to reduce the problem to the integration of a second order ODE with a parameter to be determined. Therefore with standard numerical methods we are able to find candidates for stable isoperimetric surfaces, calculate their area and compare with the inequalities \eqref{desProbada} and \eqref{desConjetura}.

The article is organized as follows. In Section \ref{sec2} we describe the family of spacetimes found in \cite{Bonnor98}. Then, in Section \ref{sec3}, we calculate the ODE satisfied by CMC surfaces and discuss the particularities of its integration and the numerical scheme. The results on the relation between area an charge are discussed in Section \ref{sec4} This leads to an interesting limit manifold that is analyzed in Section \ref{sec5}. Finally the conclusions are presented in Section \ref{sec6}.

\section{The metric for the ECD prolate spheroid}\label{sec2}

In \cite{Bonnor98} Bonnor presents a family of static, axially symmetric, asymptotically flat solutions of the Einstein-Maxwell field equations. The matter content of the spacetimes corresponds to a prolate spheroid of ECD surrounded by vacuum. The metric in prolate spheroidal coordinates is
\begin{equation}\label{metric4}
 ds^2=-U^{-2}dt^2+a^2U^2\big(X(du^2+d\theta^2)+\sinh^2u\cos^2\theta d\phi^2\big),
\end{equation}
where $a$ is a constant and
\begin{equation}
 X=\cosh^2u-\sin^2\theta.
\end{equation}
The boundary between the matter and vacuum regions is defined by choosing a positive constant $u_0$, with $u<u_0$ corresponding to the region where the matter is, and $u>u_0$ corresponding to the vacuum region. The metric function $U$ is
\begin{equation}
 U = \left\{ \begin{array}{l l}
              1+\frac{m}{a}\ln\coth\frac{u}{2}, & u > u_0,\\
              1+\frac{m}{a}\Big(\ln\coth\frac{u_0}{2}+\frac{u_0^5-u^5}{5u_0^4\sinh u_0}\Big), & u < u_0.
             \end{array}\right.
\end{equation}
We refer the reader to \cite{Bonnor98} for details regarding the field equations. The parameter $m$ is the ADM mass of the spacetime, and as we are dealing with ECD it is also the total electric charge, $Q=m$. The parameter $a$ comes from the definition of prolate spheroidal coordinates, as in fact such coordinates are not a single system of coordinates but a family. It is interesting that changing this parameter changes the spacetime, as it changes the coordinates on which the spacetime is constructed.

In \cite{Bonnor98} the geometric properties of the ECD spheroid are discussed. If we consider $u_0$ as a parameter defining a family of spacetimes, while the other parameters are kept constant, then it is shown that the equatorial perimeter and area of the spheroid go to zero as $u_0$ goes to zero, while the polar perimeter diverges. This implies that given any positive number $k$, there is always possible to choose $u_0$ such that
\begin{equation}
 A_{u_0} < k\,m^2,
\end{equation}
or in terms of the total charge
\begin{equation}\label{ineqFail}
 A_{u_0} < k\,Q^2,
\end{equation}
where $A_{u_0}$ is the area of said spheroid. This is used in \cite{Bonnor98} as a counterexample to the conjecture that any mass distribution whose area is small enough with respect to its mass should form a black hole.

\section{CMC surfaces}\label{sec3}

As the maximal initial hypersurface we simply take a $t=constant$ slice. The extrinsic curvature vanishes and the induced metric is
\begin{equation}
 ds^2=a^2U^2\big(X(du^2+d\theta^2)+\sinh^2u\cos^2\theta \,d\phi^2\big).
\end{equation}
If we define the parameter
\begin{equation}
 \mu = \frac{m}{a}
\end{equation}
we see that $a^2$ is just an overall constant factor in the metric, which in fact is also an overall factor in the spacetime metric if we rescale the coordinate $t$. Therefore we concern ourselves with the metric
\begin{equation}\label{metric3}
 ds^2=U^2\big(X(du^2+d\theta^2)+\sinh^2u\cos^2\theta \,d\phi^2\big)
\end{equation}
and powers of $a$ can be restated later whenever necessary. As we only need to consider the vacuum region or the surface of the spheroid we have
\begin{equation}
 U=1+\mu\ln\coth\frac{u}{2}.
\end{equation}

In order to see why the inequality \eqref{ineqFail} is possible, we calculate the mean extrinsic curvature, $\chi$, of the $u=constant$ spheroids. A lengthy but straightforward computation gives
\begin{equation}
 \chi = \frac{U\cosh u (2\sinh^2u+\cos^2\theta)-2\mu X}{U^2X^\frac{3}{2}\sinh u}.\label{chispheroid}
\end{equation}
This shows explicitly that the $u_0$ spheroid is not an isoperimetric surface. We indirectly knew this because such spheroids can be made to violate the inequality \eqref{desProbada}.

As noted, a necessary condition for a surface to be isoperimetric is that its mean extrinsic curvature has to be constant. We describe a generic axisymmetric surface by the embedding $(u,\theta,\phi)=(s(\theta),\theta,\phi)$, where $s(\theta)$ is a function of $\theta$ only and we have made an abuse of notation by denoting also by $\theta$ and $\phi$ the coordinates on the surface.
The mean extrinsic curvature for such a surface is
\begin{eqnarray}\label{eqMean}
 \chi & = & -\frac{1}{U\sqrt{X(1+s'^2)}}\Bigg[\frac{s''}{1+s'^2}-\frac{1}{2}s'\sin(2\theta)\Big(\frac{1}{X}+\frac{1}{\cos^2\theta}\Big) \\
 && -\frac{1}{2}\sinh(2s)\Big(\frac{1}{X}+\frac{1}{\sinh^2s}\Big)+\frac{2\mu}{U\sinh s}\Bigg],
\end{eqnarray}
where
\begin{equation}
 X = \cosh^2 s-\sin^2 \theta, \qquad U = 1+\mu\ln\coth\frac{s}{2},
\end{equation}
and prime denotes derivative with respect to $\theta$.
In the special case $s(\theta)=constant$ we recover the formula for the spheroids \eqref{chispheroid}. Equation \eqref{eqMean} can be written in a more suggestive form as a differential equation for $s$, if we consider $\chi$ as a constant parameter,
\begin{eqnarray}\label{isSurfEq}
 s'' & = & (1+s'^2)\Bigg[-\chi U \sqrt{X(1+s'^2)}+\frac{1}{2}s'\sin(2\theta)\Big(\frac{1}{X}+\frac{1}{\cos^2\theta}\Big) \\
 && +\frac{1}{2}\sinh(2s)\Big(\frac{1}{X}+\frac{1}{\sinh^2 s}\Big)-\frac{2\mu}{U\sinh s}\Bigg].
\end{eqnarray}
A solution of \eqref{isSurfEq} corresponds to a CMC surface and hence we want to integrate it from $\theta=-\frac{\pi}{2}$ to $\theta=0$. The first thing to consider is that the solution needs to represent a smooth enough surface and this means that at least $s'\left(-\frac{\pi}{2}\right)=s'(0)=0$. Also, the differential equation \eqref{isSurfEq} is singular at $\theta=-\frac{\pi}{2}$, that is, the r.h.s. formally diverges at that point. This is a coordinate problem due to the spheroidal coordinates being singular there. This issue can be remedied, as discussed below, and the r.h.s. has a well defined limit if $s'\left(-\frac{\pi}{2}\right)=0$, coincident with one of the previous conditions.

Due to the complicated nature of \eqref{isSurfEq} we integrate it numerically using a shooting method. We fix the initial conditions $s_0=s\left(-\frac{\pi}{2}\right)$ and $s_1=s'\left(-\frac{\pi}{2}\right)=0$, and make a guess for $\chi$. We can then integrate the differential equation. We use as error function the value of $s'$ at the end point of the integration, that is, $s'(0)$. We update the guess for $\chi$ using the Newton-Raphson method,
\begin{equation}
 \chi_{k+1}=\chi_{k}-\frac{s'(0)}{\dot{s}'(0)},
\end{equation}
where dot means derivative with respect to $\chi$. We iterate until the value of the error function is small enough. After this process we obtain the function $s(\theta)$ and the corresponding $\chi$. By varying $s_0$ we change the surface found and by varying $\mu$ in \eqref{isSurfEq} we change the metric in which the surface is embedded, correspondingly we denote by $s_{s_0,\mu}$ the solution and by $\chi_{s_0,\mu}$ its mean extrinsic curvature.

As said, the differential equation \eqref{isSurfEq} is singular at $\theta=-\frac{\pi}{2}$. Although the limit is well defined, the singularity poses problems for the numerical integration. In order to circumvent this issue, before integrating we extrapolate the initial conditions from $\theta=-\frac{\pi}{2}$ to $\theta=-\frac{\pi}{2}+\delta$, where $\delta$ is the size of the first step on the integration grid. The extrapolation is done via a Taylor expansion to second order
\begin{equation}
 s(\theta) \approx s_0+s_1\left(\theta+\frac{\pi}{2}\right)+\frac{1}{2}s_2\left(\theta+\frac{\pi}{2}\right)^2.
\end{equation}
The coefficients $s_0$ and $s_1$ could in principle be prescribed, but due to \eqref{isSurfEq} being singular only $s_0$ can be freely prescribed. In order for \eqref{isSurfEq} to have a well defined limit at $\theta=-\frac{\pi}{2}$ we need
\begin{equation}
 s_1=0.
\end{equation}
Once this is ensured, the r.h.s. of \eqref{isSurfEq} has a well defined limit that depends on $s_2$, and equating this with the l.h.s., i.e. $s_2$, we obtain
\begin{equation}
 s_2=-\frac{1}{2}\chi U_0 \sinh s_0 + \coth s_0 -\frac{\mu}{U_0\sinh s_0},
\end{equation}
where
\begin{equation}
 U_0 = 1+\mu\ln\coth\frac{s_0}{2}.
\end{equation}
Summarizing, the new initial conditions are
\begin{equation}
 s\left(-\frac{\pi}{2}+\delta\right) = s_0 + \frac{1}{2}s_2\delta^2, \qquad s'\left(-\frac{\pi}{2}+\delta\right) = s_2\delta,
\end{equation}
and the integration is performed from $\theta=-\frac{\pi}{2}+\delta$ to $\theta=0$.

An accompanying strategy to help with the singular limit is to use a grid with unevenly spaced points, in order to have more resolution close to the singular end. After some trials we decided to use a grid where the point $i$ is at position
\begin{equation}
 \theta_i = -\frac{\pi}{2}+\frac{\pi}{2}\left(\frac{i}{n}\right)^{1.4},\quad 0\leq i \leq n.
\end{equation}

\begin{figure}
  \centering
  {{}}\def\svgwidth{.8\textwidth}
  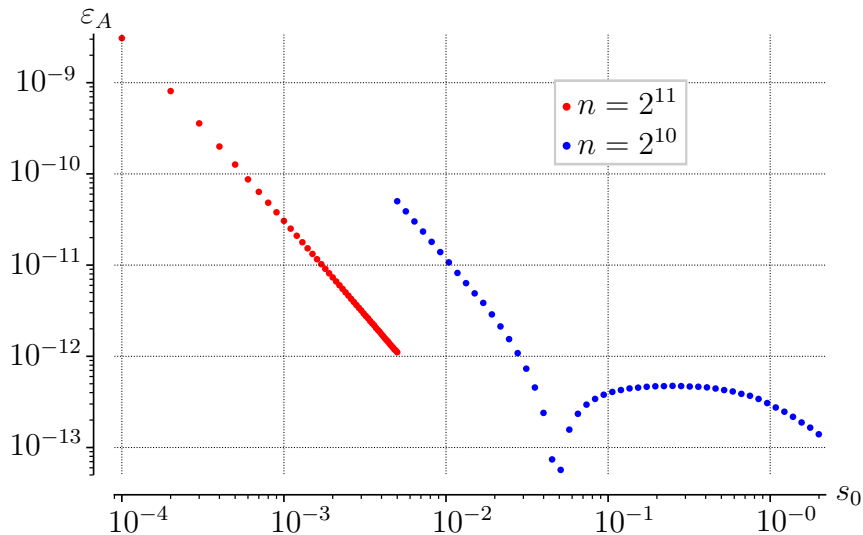
  \caption{Relative error in the area for the Euclidean case.}
  \label{fig0}
\end{figure}

Once we have obtained $s_{s_0,\mu}$ we calculate its area, which is given by
\begin{equation}
 A_{s_0,\mu} = 4\pi\int_{-\frac{\pi}{2}}^{0}U^2\sqrt{X(1+s'^2)}\sinh s\cos{\theta}\,d{\theta}.
\end{equation}
To perform the ODE integration we use the Runge-Kutta-Fehlberg method, and for the area integration the  composite Simpson's rule. Both methods are already implemented in SageMath \cite{SageMath}. To test the accuracy of the numerical solution we use the $\mu=0$ case, which is simply Euclidean space and where we know that the isoperimetric surfaces are spheres, which in prolate spheroidal coordinates have the expression
\begin{equation}
 s_E(\theta)=\mbox{arccosh}\sqrt{\cosh^2 s_0+\cos^2\theta},
\end{equation}
and for which
\begin{equation}
 \chi_E = \frac{2}{\cosh s_0},\qquad A_E = 4\pi\cosh^2(s_0).
\end{equation}
Comparing the numerical solution to the exact solution we decided to use a grid with $n=2^{10}$ points for $s_0> 0.005$ and with $n=2^{11}$ points for $s_0\leq 0.005$ which ensures enough accuracy. The relative error in the area, which is our main concern in the numerical scheme, is shown in Figure \ref{fig0}.

\begin{figure}
  \centering
  {{}}\def\svgwidth{\textwidth}
  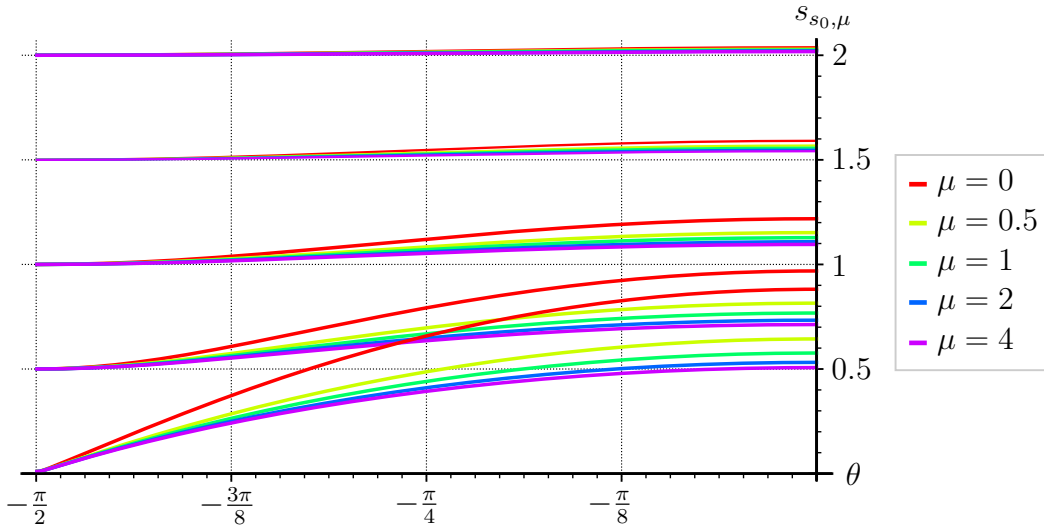
  \caption{CMC surfaces.}
  \label{fig1}
\end{figure}

In Figure \ref{fig1} we have plotted several solutions to \eqref{isSurfEq} with varying $s_0$ and $\mu$, which illustrates the general tendency of the CMC surfaces. As expected from the $\mu=0$ case, each $s_{s_0,\mu}(\theta)$ is an increasing function of $\theta$. Also expected, if we fix $\mu$, then $s_{s_0,\mu}(\theta)$ is an increasing function of $s_0$ for every $\theta$ and therefore the solution $s_{s_0=0,\mu}(\theta)$ bounds from below all the other solutions. Finally, if we fix $s_0$, then $s_{s_0,\mu}(\theta)$ is a decreasing function of $\mu$, and the solution does not go to zero but there exists a limit solution as $\mu\rightarrow\infty$.

\begin{figure}
  \centering
  {{}}\def\svgwidth{.8\textwidth}
  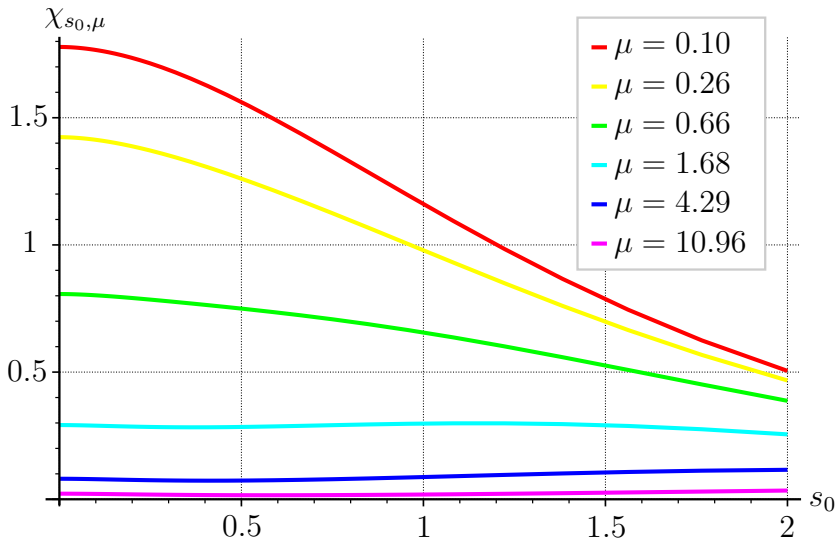
  \caption{Mean extrinsic curvature.}
  \label{fig2}
\end{figure}

In Figure \ref{fig2} we show the mean extrinsic curvature of the CMC surfaces as a function of $s_0$ for several values of $\mu$. In general, as expected, $\chi_{s_0,\mu}$ is a decreasing function of $s_0$, although for some values of $\mu$ there is a range in $s_0$ where $\chi_{s_0,\mu}$ is an increasing function. In all cases the mean extrinsic curvature goes to zero as $s_0\rightarrow\infty$. Also, if we fix $s_0$, then $\chi_{s_0,\mu}$ is a decreasing function of $\mu$ and goes to $0$ as $\mu\rightarrow\infty$.

\section{Relation between area and charge}\label{sec4}

Our main interest in this work is to analyze the relation between area and charge for the surfaces we have integrated, and compare with the inequalities \eqref{desProbada} and \eqref{desConjetura}. All the surfaces we consider enclose the ECD spheroid and therefore their charge is the charge of the spacetime. To make the comparison easier we define the quantity
\begin{equation}
 q_{s_0,\mu} = \frac{A_{s_0,\mu}}{4\pi\mu^2}.
\end{equation}
This quotient is the same as $A/(4\pi Q^2)$ once we reinstate the factors of $a$. We know that $q_{s_0,\mu}$ is not bounded above, as the spacetime is asymptotically flat and then $A_{s_0,\mu}$ can be made as large as we want by taking $s_0$ big enough. On the other hand, $q_{s_0,\mu}$ has to be bounded below away from zero if $s_{s_0,\mu}$ is a stable isoperimetric surface. In order to compare with the inequalities \eqref{desProbada} and \eqref{desConjetura} we note that if they are satisfied by $s_{s_0,\mu}$ then $q_{s_0,\mu}\geq k$ with $k=\frac{1}{3}$ for \eqref{desProbada} and $k=\frac{4}{9}$ for \eqref{desConjetura}.

\begin{figure}
  \centering
  {{}}\def\svgwidth{.8\textwidth}
  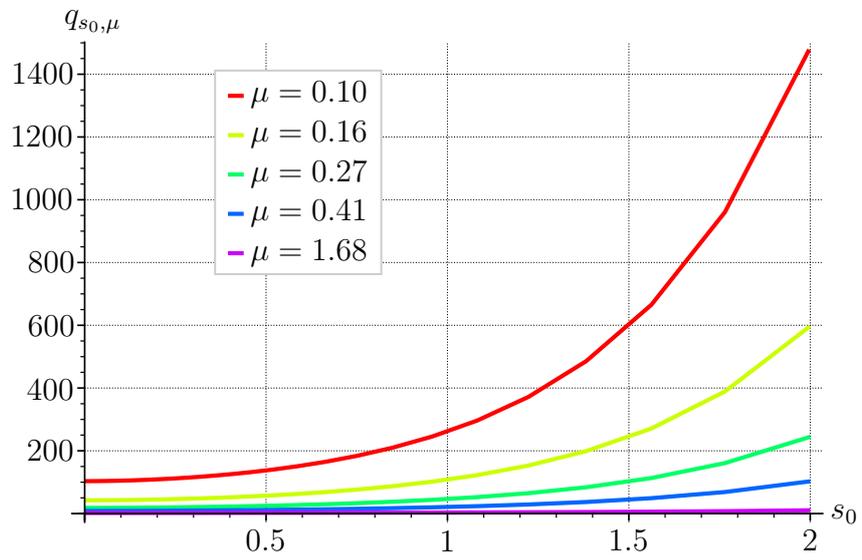
  \caption{Area-charge quotient.}
  \label{fig3}
\end{figure}

\begin{figure}
  \centering
  {{}}\def\svgwidth{.8\textwidth}
  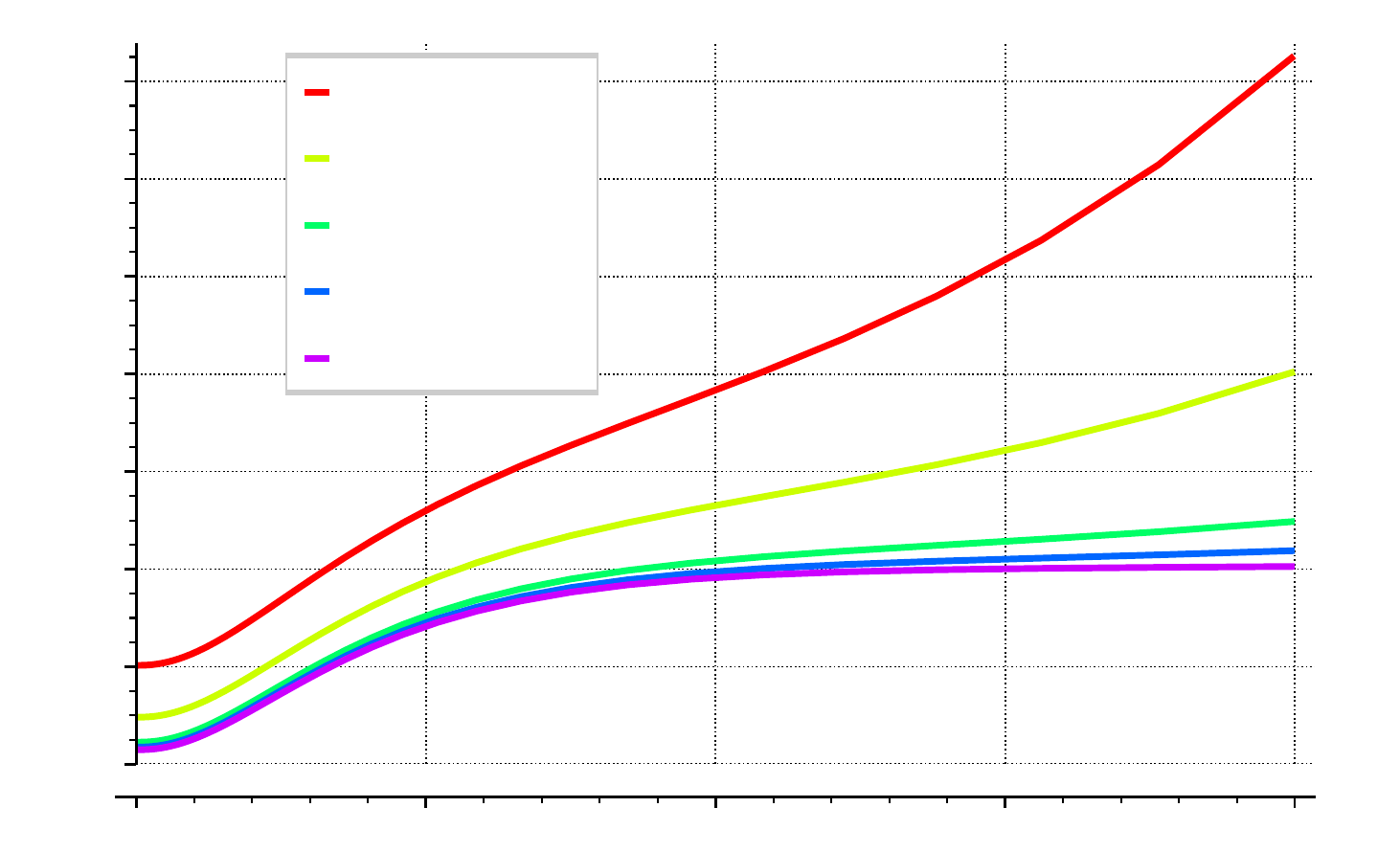
  \caption{Area-charge quotient for large $\mu$.}
  \label{fig4}
\end{figure}

\begin{figure}
  \centering
  {{}}\def\svgwidth{.8\textwidth}
  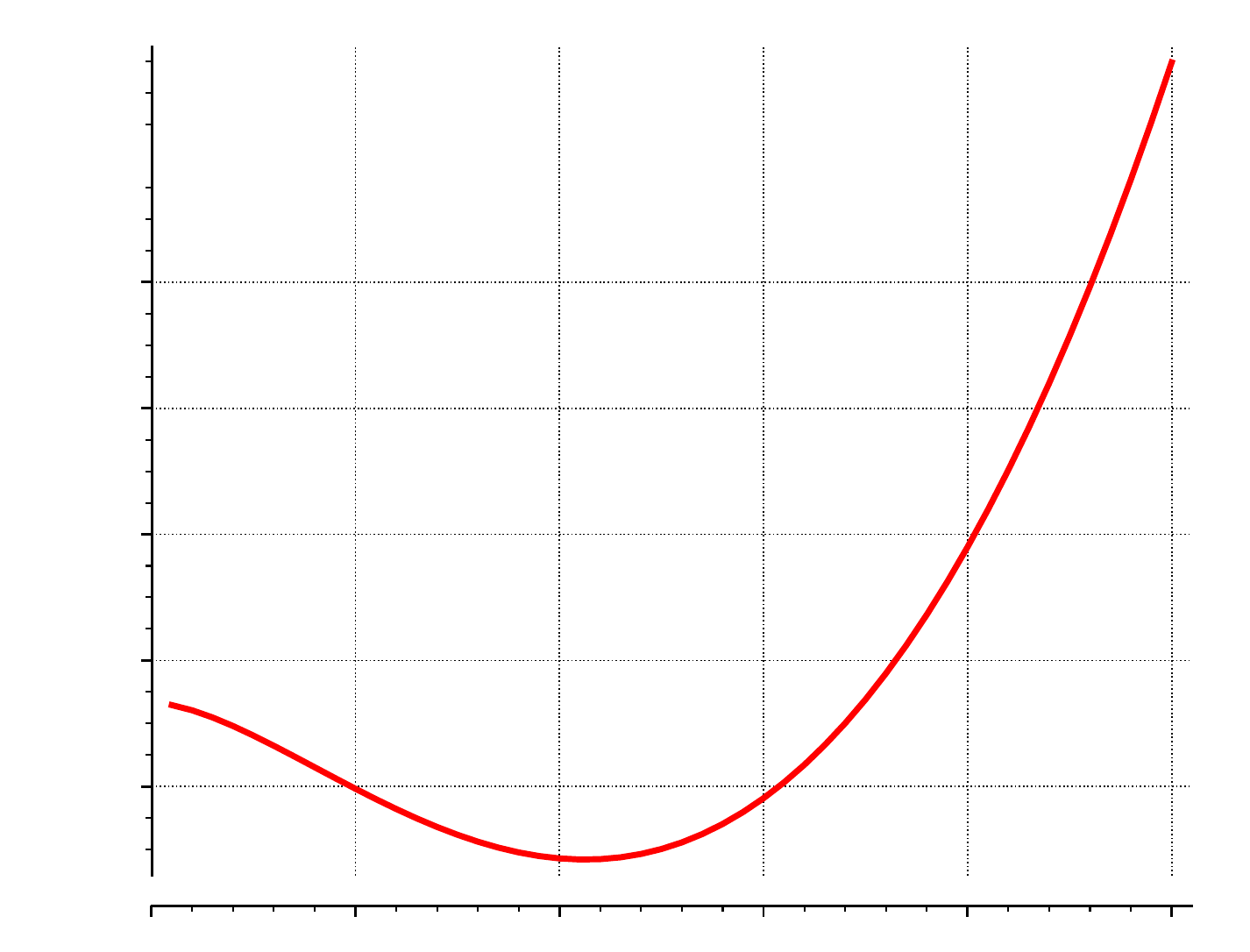
  \caption{Area-charge quotient for small $s_0$.}
  \label{fig5}
\end{figure}

In Figure \ref{fig3} we have plotted $q_{s_0,\mu}$ as a function of $s_0$ for several values of $\mu$. We see that $q_{s_0,\mu}$ is an increasing function of $s_0$, as expected, and that there is a well defined limit for $q_{s_0,\mu}$ as $s_0$ goes to zero, which can be quite large if $\mu$ is small. We plot again $q_{s_0,\mu}$ as a function of $s_0$ in Figure \ref{fig4}, but for higher values of $\mu$. We observe the same behavior as before, although the curves get closer as $\mu$ increases and there is a limiting curve. We see already that the minimum of $q_{s_0,\mu}$ seems to be above $0.96$, a value that is far from $\frac{1}{3}$ or $\frac{4}{9}$.  There is a subtlety regarding the case when $s_0$ is close to zero. In Figure \ref{fig5} we have zoomed into the region close to $s_0=0$ for $\mu=28$. We see that the minimum of $q_{s_0,\mu}$ is not attained at $s_0=0$ but at around $s_0=0.0021$. Analyzing the results of the integrations we conclude that such minimum first appears for $\mu\approx 1$, although we have not tried to find exactly when this happens. Said minimum moves to the right and settles at around $s_0=0.0026$ with increasing $\mu$.

\begin{figure}
  \centering
  {{}}\def\svgwidth{.8\textwidth}
  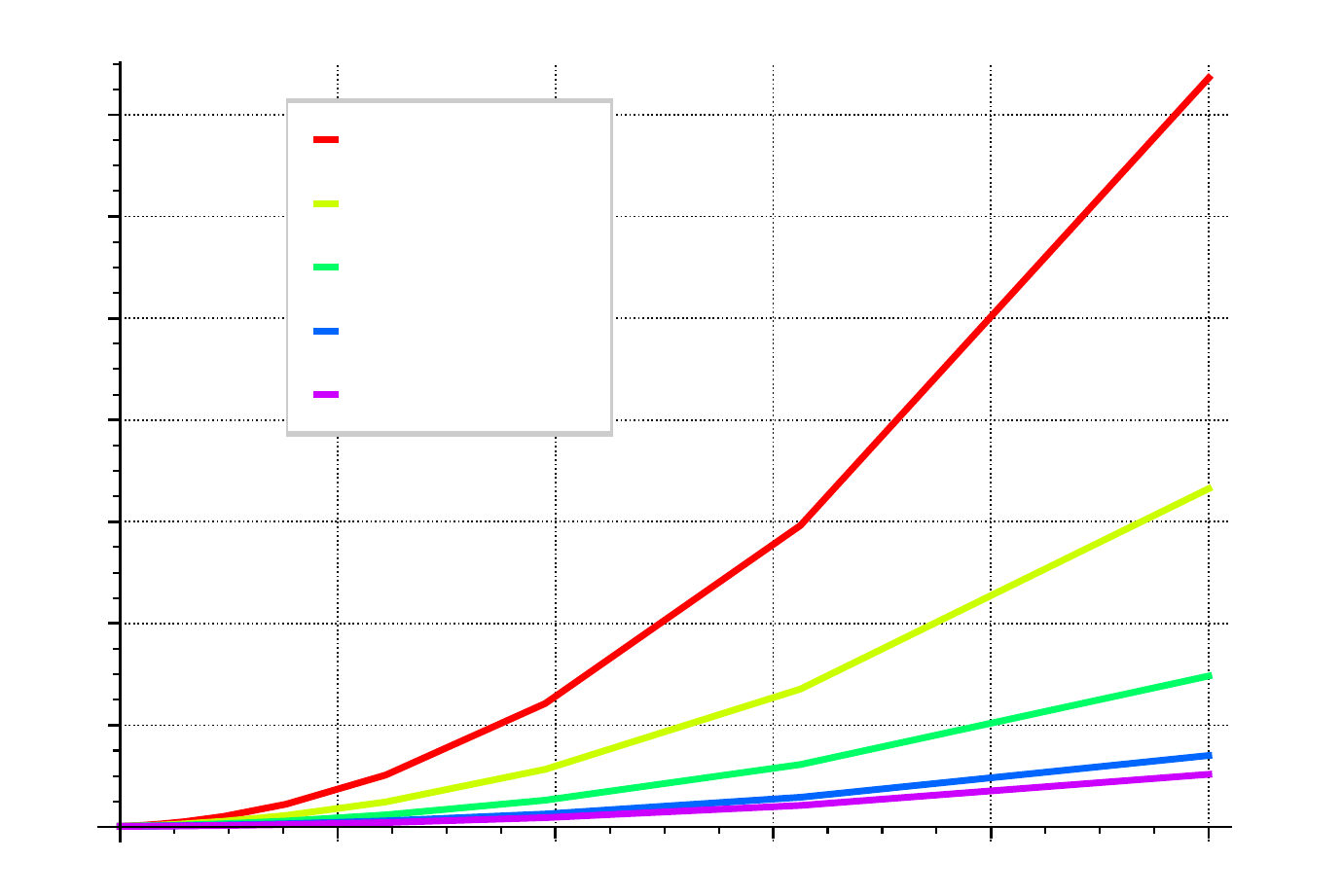
  \caption{Area-charge quotient.}
  \label{fig6}
\end{figure}

\begin{figure}
  \centering
  {{}}\def\svgwidth{.8\textwidth}
  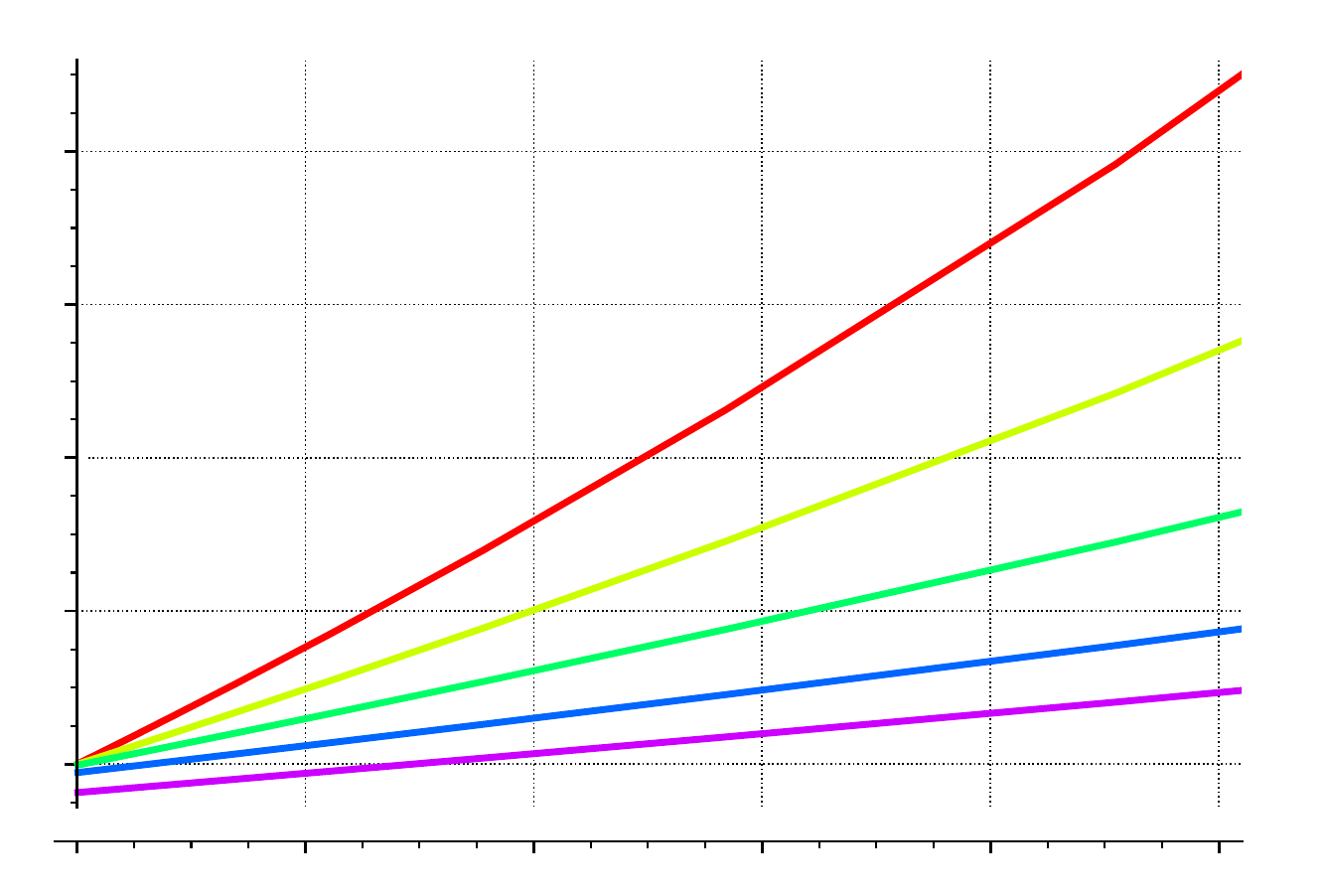
  \caption{Area-charge quotient for large $\mu$.}
  \label{fig7}
\end{figure}

From the plots we note that $q_{s_0,\mu}$ is a decreasing function of $\mu$ if one keeps $s_0$ constant. To analyze this we show in Figure \ref{fig6} $q_{s_0,\mu}$ as a function of $\mu^{-1}$, which highlights its behavior for large $\mu$. We see that it is an increasing function of $\mu^{-1}$. To see the limit as $\mu^{-1}\rightarrow 0$ we zoom in in Figure \ref{fig7}, showing that $q_{s_0,\mu}$ has a well defined limit there and that said limit does depend on $s_0$. As already noted we have $q_{s_0,\mu}>0.96$. The behavior of $q_{s_0,\mu}$ suggest taking the limit $\mu\rightarrow\infty$, which we do in the following section.

\begin{figure}
  \centering
  {{}}\def\svgwidth{\textwidth}
  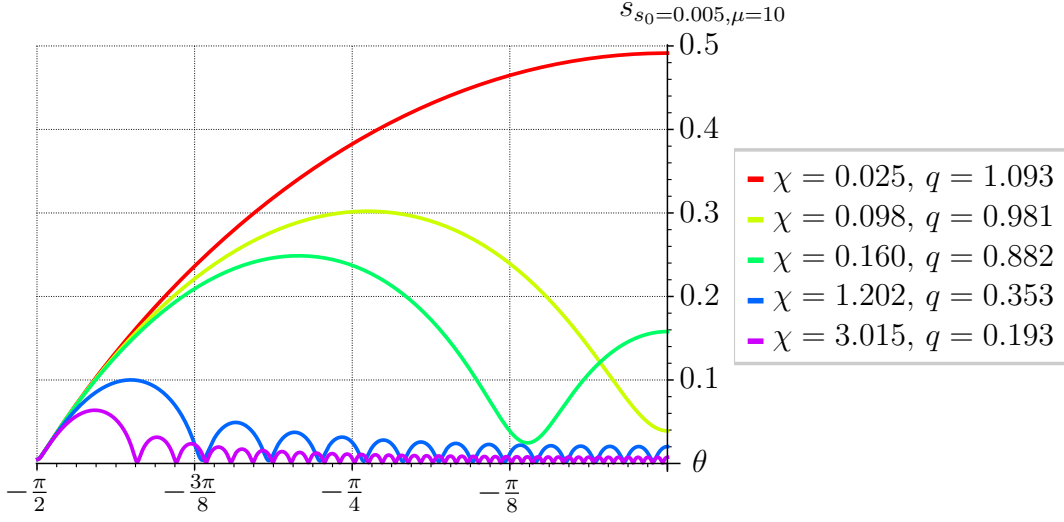
  \caption{Other CMC surfaces.}
  \label{fig11}
\end{figure}

To close this section we point out that there are other CMC surfaces besides those considered so far. In Figure \ref{fig11} we have plotted five CMC surfaces, all with $s_0=0.005$ and for $\mu=10$. The one in red corresponds to a candidate for a stable isoperimetric surface, while the other four are not candidates, as they are concave and therefore they are not stable isoperimetric surfaces. It is interesting that these surfaces can violate both \eqref{desConjetura} and \eqref{desProbada}, which again shows that the stability requirement is crucial.

\section{The limit $\mu\rightarrow\infty$}\label{sec5}

From the results so far it is clear that the lowest possible value for $q_{s_0,\mu}$ is attained in the limit $\mu\rightarrow \infty$ and for a CMC surface with $s_0$ close to zero. It is quite interesting that the isoperimetric problem can be analyzed in the limit $\mu\rightarrow\infty$ as we show in the present section. We start by defining a limit metric,
\begin{equation}
 d\hat{s}^2 = \lim_{\mu\rightarrow\infty} \frac{ds^2}{\mu^2} = \hat{U}^2 \left(X(du^2+d\theta^2)+\sinh^2u \cos^2\theta d\phi^2\right),
\end{equation}
where
\begin{equation}
 \hat{U} = \ln\coth\frac{u}{2}.
\end{equation}
The quantities that we calculated for $ds^2$ can be obtained for $d\hat{s}^2$ by performing the analogous calculations or simply by multiplying the adequate factors of $\mu$ and then taking the limit. In particular, the equation corresponding to a CMC surface is
\begin{eqnarray}
 s'' & = & (1+s'^2)\Bigg[-\hat{\chi} \hat{U} \sqrt{X(1+s'^2)}+\frac{1}{2}s'\sin(2\theta)\Big(\frac{1}{X}+\frac{1}{\cos^2\theta}\Big) \\
 && +\frac{1}{2}\sinh(2s)\Big(\frac{1}{X}+\frac{1}{\sinh^2 s}\Big)-\frac{2}{\hat{U}\sinh s}\Bigg],
\end{eqnarray}
where
\begin{equation}
 \hat{\chi}=\lim_{\mu\rightarrow\infty}\mu\chi.
\end{equation}
For the numerical integration, the Taylor approximation coefficients are
\begin{equation}
 s_1 = 0,\qquad s_2=-\frac{1}{2}\hat{\chi} \hat{U}_0 \sinh s_0 + \coth s_0 -\frac{1}{\hat{U}_0\sinh s_0}.
\end{equation}
Also
\begin{equation}
 \hat{A}_{s_0} = \lim_{\mu\rightarrow\infty}\frac{A_{s_0,\mu}}{\mu^2}  = 4\pi\int_{-\frac{\pi}{2}}^{0}\hat{U}^2\sqrt{X(1+s'^2)}\sinh s\cos{\theta}\,d{\theta},
\end{equation}
and the quotient between area and charge takes the form
\begin{equation}
 \hat{q}_{s_0} = \frac{\hat{A}_{s_0}}{4\pi} = \lim_{\mu\rightarrow\infty}q_{s_0,\mu}.
\end{equation}

It is worth noticing that the limit metric can be written in spherical coordinates as
\begin{equation}
 d\hat{s}^2 = \hat{U}^2(dr^2+r^2 d\Omega^2).
\end{equation}
If we consider $r$ large, then
\begin{equation}
 d\hat{s}^2 \approx \frac{dr^2}{r^2} + d\Omega^2,
\end{equation}
with the relative error in the metric functions being of order $r^{-1}$. If we define $\hat{r} = \ln{r}$ then
\begin{equation}
 d\hat{s}^2 \approx d\hat{r}^2 + d\Omega^2,
\end{equation}
which shows that the metric is asymptotically cylindrical and that for $r$ large the surfaces $r=constant$ are approximate isoperimetric surfaces, with area $\hat{A}_{s_0\ra\infty}= 4\pi$, which gives $\hat{q}_{s_0\ra\infty} = 1$.

\begin{figure}
  \centering
  {{}}\def\svgwidth{.8\textwidth}
  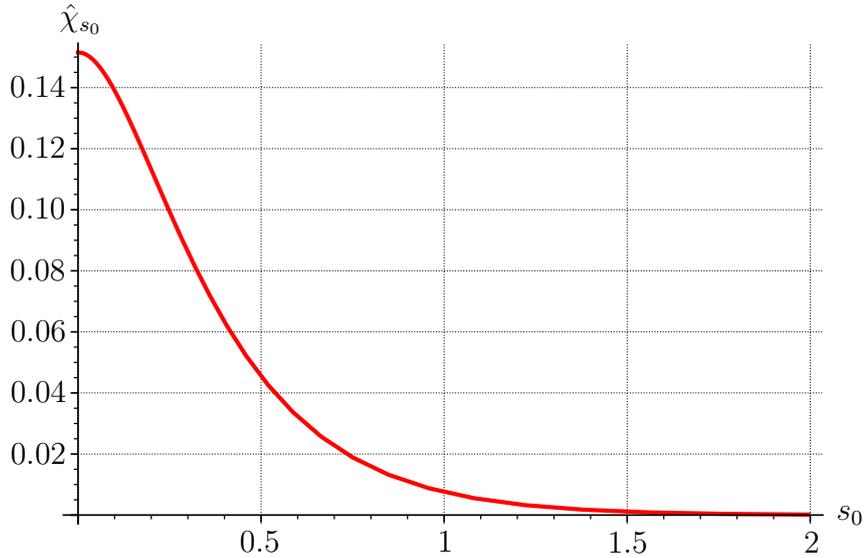
  \caption{Mean extrinsic curvature for $\mu\rightarrow\infty$.}
  \label{fig8}
\end{figure}

\begin{figure}
  \centering
  {{}}\def\svgwidth{.8\textwidth}
  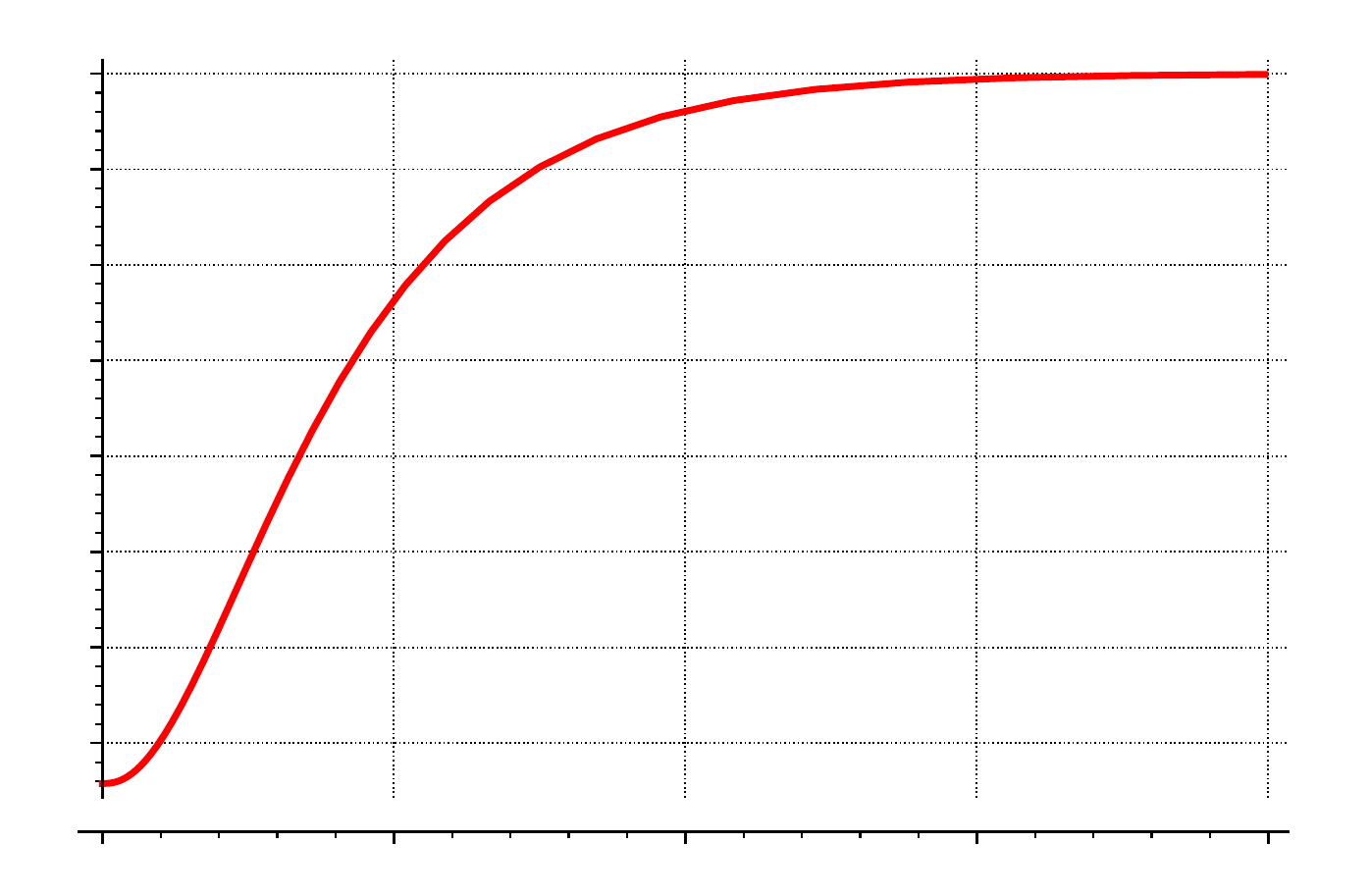
  \caption{Area-charge quotient for $\mu\rightarrow\infty$.}
  \label{fig9}
\end{figure}

\begin{figure}
  \centering
  {{}}\def\svgwidth{.8\textwidth}
  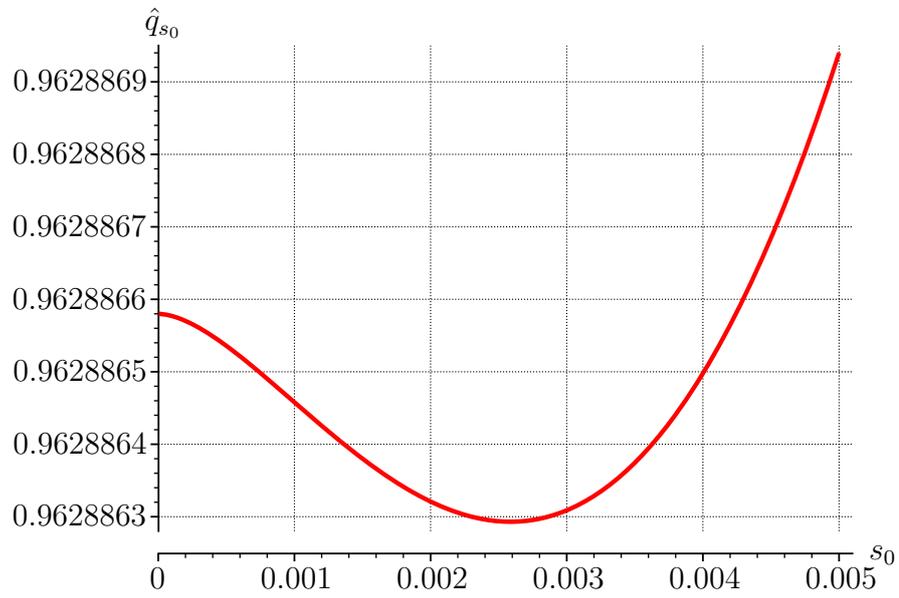
  \caption{Area-charge quotient for $\mu\rightarrow\infty$ close to $s_0=0$.}
  \label{fig10}
\end{figure}

For the integrations corresponding to this section we increased the number of grid points to $n=2^{12}$ for $s_0\leq0.005$, in order to have better accuracy. From the numerical results, we first plot $\hat{\chi}_{s_0}$ in Figure \ref{fig8}. We see that it is a decreasing function that goes to zero. To analyze the area-charge quotient, in Figure \ref{fig9} we have plotted $\hat{q}_{s_0}$, and we zoom in close to $s_0=0$ in Figure \ref{fig10}. We find that the minimum of $\hat{q}_{s_0}$ is attained at $s_0=0.00259$ and with a value $\hat{q}_{s_0}=0.9628862930603$.
From this we conclude that all CMC surfaces that are candidates to be stable isoperimetric surfaces satisfy
\begin{equation}
 q_{s_0,\mu} \geq 0.9628862930603,
\end{equation}
which as noted before is sufficiently above $\frac{1}{3}$ and $\frac{4}{9}$ as not to present a challenge for the inequalities \eqref{desProbada} and \eqref{desConjetura}.

\section{Conclusions}\label{sec6}

We have studied the spacetime found by Bonnor \cite{Bonnor98} in the light of the geometrical inequality \eqref{desProbada}. Through a shooting method we have obtained numerically a foliation of part of the maximal initial slice by CMC surfaces which are candidates to be stable isoperimetric surfaces. By the results in \cite{HuiskenYau96}, \cite{EichmairMetzger2013a} and \cite{EichmairMetzger2013b} we know that for $s_0$ large enough they are indeed stable isoperimetric surfaces, and although we do not know how large $s_0$ needs to be it seems that the foliation gets quite close to the ECD spheroid even when $\mu$ diverges.

With these CMC surfaces we tested the inequalities \eqref{desProbada} and \eqref{desConjetura}, showing that for the family of spacetimes
\begin{equation}\label{desECD}
 q_{s_0,\mu} \geq 0.9628862930603.
\end{equation}
So the bound on $q_{s_0,\mu}$ is far from $\frac{1}{3}$ and $\frac{4}{9}$, and therefore does not present a challenge to the geometrical inequalities. This not being close to saturate the inequalities can be interpreted as the spacetime being far from extremality in terms of its charge density. An interesting consequence of searching for \eqref{desECD} is that we were led naturally to consider a limit metric, which instead of an asymptotically flat end has a cylindrical end, and were the minimum of \eqref{desECD} is attained.

We also found CMC surfaces that are not stable isoperimetric, showing that some of them do not satisfy \eqref{desProbada} or \eqref{desConjetura}, illustrating that the stability requirement is fundamental for the inequalities to be valid.

We have not analyzed several interesting properties of the surfaces, most notably their stability, but it is a complicated problem that did not add to  the main objective of the work, as the inequalities \eqref{desProbada} and \eqref{desConjetura} were far from being challenged. Also, we did not investigate the reasons why the minimum of $q_{s_0,\mu}$ is achieved for a positive value of $s_0$ for $\mu$ big enough, the increasing value of $\chi_{s_0,\mu}$ as a function of $s_0$, or the relation between the isoperimetric surfaces and the spacetime, if there is any. Finally, and far more reaching, is the question of finding a good measure of the ``size'' of an object. We have not tried to analyze other proposed measures of size and compare to the isoperimetric surfaces.

\section{Acknowledgments}

I would like to dedicate this article to the memory of Sergio Dain, he suggested this problem, generous as he was with ideas, and discussions with him were always enlightening.

The numerical computations were performed and the figures produced in SageMath \cite{SageMath}.

\end{document}